\documentclass[conference]{IEEEtran}
\IEEEoverridecommandlockouts
\usepackage[noadjust]{cite}
\usepackage{amsmath,amssymb,amsfonts,amsthm}
\usepackage{graphicx}
\usepackage{booktabs}
\usepackage{algorithm}
\usepackage{mathrsfs}
\usepackage[hyphens]{url}
\usepackage[hidelinks]{hyperref}
\hypersetup{breaklinks=true}

\usepackage{array}
\usepackage{ragged2e}
\newcolumntype{L}[1]{>{\RaggedRight\hspace{0pt}}p{#1}}
\newcolumntype{C}[1]{>{\centering\let\newline\\\arraybackslash\hspace{0pt}}m{#1}}

\usepackage{tikz}
\usetikzlibrary{arrows,shapes,intersections,decorations,plotmarks,math}

\def\BibTeX{{\rm B\kern-.05em{\sc i\kern-.025em b}\kern-.08em
    T\kern-.1667em\lower.7ex\hbox{E}\kern-.125emX}}

\begin{document}

\title{Explainable AI based System for Supply Air Temperature Forecast \\
}

\author{\IEEEauthorblockN{Marika Eik}
\IEEEauthorblockA{\textit{EEUAS, R8Technologies, TalTech}\\
Tallinn, Estonia \\
marika.eik@r8tech.io}
\and
\IEEEauthorblockN{Ahmet Kose}
\IEEEauthorblockA{\textit{R8Technologies, TalTech}\\
Tallinn, Estonia \\
ahmet.kose@r8tech.io}
\and
\IEEEauthorblockN{Hossein Nourollahi Hokmabad, Juri~Belikov}
\IEEEauthorblockA{\textit{Tallinn University of Technology,}\\
Tallinn, Estonia \\
\{hossein.nourollahi, juri.belikov\}@taltech.ee}
}

\maketitle
\begin{abstract}
This paper explores the application of Explainable AI (XAI) techniques to improve the transparency and understanding of predictive models in control of automated supply air temperature (ASAT) of Air Handling Unit (AHU). The study focuses on forecasting of ASAT using a linear regression with Huber loss. However, having only a control curve without semantic and/or physical explanation is often not enough. The present study employs one of the XAI methods: Shapley values, which allows to reveal the reasoning and highlight the contribution of each feature to the final ASAT forecast. In comparison to other XAI methods, Shapley values have solid mathematical background, resulting in interpretation transparency. The study demonstrates the contrastive explanations--slices, for each control value of ASAT, which makes it possible to give the client objective justifications for curve changes.
\end{abstract}

\section{Introduction}
Artificial Intelligence (AI) solutions are rapidly expanding and increasingly integrating with conventional systems. However, as the complexity of these models grows—particularly deep learning (DL) models with billions of parameters—their reasoning and outcomes are becoming difficult for humans to interpret and justify \cite{MeasMKTLLPB2022}. This opaque decision-making process limits the applicability of this promising technology in high-risk fields, such as the aviation industry, healthcare, military, and very large-scale systems like electricity grid, where any error or unanticipated risk could result in unpredictable consequences. 

To mitigate concerns and reduce risk factors associated with AI models, researchers have introduced Explainable AI (XAI) methods. These techniques bridge the gap between AI-generated outcomes and expert assessments, helping to prevent unintended negative consequences. XAI focuses on creating tools that make black-box models more transparent, clarify complex systems, and improve human reasoning \cite{BIPARVA20232383}. In this context, two key concepts are interpretability and explainability. Interpretability refers to how easily humans can understand a model's predictions based solely on its structure and design, while explainability focuses on how well humans can reason about the factors influencing the model's predictions \cite{BAUR2024100358}.

Techniques such as SHapley Additive exPlanations (SHAP) \cite{NIPS2017_7062} and Local Interpretable Model-Agnostic Explanations (LIME) \cite{ribeiro2016should}, aim to clarify the contribution of each feature to the final model decisions and thus enhance the model explainability. For example, DeepLIFT approach is introduced to approximate Shapley values in an recurrent neural network (RNN)-based energy forecasting model to reveal how features at different time intervals influence neuron activation across deep layers, thereby providing explanations for the model's output forecasts \cite{SHAJALAL2024123588}. Additionally, XAI techniques can assist in feature selection by detecting and removing unnecessary features, which enhances both the model’s performance and its interpretability \cite{VANZYL2024122079}.


In commercial buildings, heating, ventilation, and air conditioning (HVAC) systems are the primary energy consumers, and optimizing their control can significantly reduce energy usage. Given that modern HVAC systems are highly complex, involving numerous sensors, actuators, and control parameters, achieving optimal performance requires careful planning and real-time adjustments. As a result, AI models have recently gained considerable attention and demonstrated promising performance in this field. These models are employed for tasks such as forecasting ambient parameters like temperature and humidity, as well as generating control commands to optimize and orchestrate HVAC operations, and diagnosing faults. However, without the aid of XAI tools, the reasoning behind these models' outcomes may be hard to understand for a human operator. This lack of transparency impedes effective human interpretation and evaluation during the control process, limiting the ability to make informed decisions or appropriately respond to issued warnings.

Recognizing this gap, this paper leverages one of the XAI techniques--Shapley values, to clarify how a regression model predicts automated supply air temperature (ASAT), a critical input for HVAC system control. By employing Shapley values-based approach, the study reveals the influence of various features on the ASAT control curve, offering insights into the physical and/or semantic reasoning behind these forecasts. The study presents contrastive explanations--slices, for each value forecasted, supporting the transparency of any changes in the control curve. The approach presented improves the understanding of modelling results from a physical point of view, providing field experts with valuable knowledge on how different features can affect the ASAT values.


\section{Technical Background and Methodology}
    

Univariate time series data are typically composed of single scalar observations that are recorded over equal timestamps. They can be modelled by typical classical methods of statistical analysis, such as Autoregressive Integrated Moving Average (ARIMA) \cite{shumway2017arima}, Seasonal ARIMA (SARIMA) \cite{li2023hybrid}, Holt-Winters \cite{chatfield1978holt}, and Kalman filtering \cite{harvey1990forecasting}. The analysis and modelling of the multivariate data, when multiple variables are recorded over time, is more complex, and classical methods are usually less effective in this case \cite{DU2020269}. 

The forecasting of time series using simpler machine learning methods may in some situations be more effective compared to the advanced ones, including neural network models \cite{brownlee.b.2018}. Work \cite{makridakis.sml.po.2018} considered RNN and LSTM methods among the less accurate ones, meaning that the complexity does not necessarily guarantee improvements in forecasting performance. To forecast time series, it is necessary to transform historical data into a supervised learning problem. This can be done by using the time steps of previous/historical observations as input values and the next time step observations as outputs. A sliding window method uses the values of the prior time steps to predict the values in the next time steps. This method is also called the \textit{lag} method. A window width or size of the lag corresponds to the number of previous time steps. A sliding window approach is the basis for how the time series data are transformed to supervised learning. A walk-forward validation method employs true observation values as input and makes the forecast.

\begin{figure}[htbp]
\centering
\includegraphics[width=0.7\columnwidth]{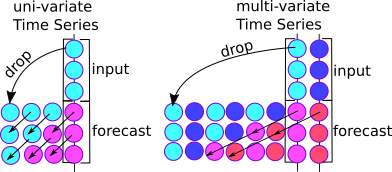}
\caption{Dropping of the time step column and shifting of the measures column.}
\label{Fig1}
\end{figure}

Figure~\ref{Fig1} illustrates the drop and shift concept. The drop value or lag, depends on the length of history that the model should consider during the learning phase. Usually, the drop value (length of the training vector) is equal to the forecast horizon.
The representation given in Fig.~\ref{Fig1} expresses a \textit{sliding window} based Hankel matrix \cite{kissel.smt.ngc.2023}, since the window of input and expected output is shifted forward in time to create new samples for the supervised learning model \cite{Hossein2023}.

\subsection{Explainablity of Linear Models}
A typical way to understand importance of features of parametric models is to consider the model coefficients--weights learned, $\beta_i$. Depending on the coefficient magnitude, representing the effect of a feature, a change in model output, while changing the feature, can be estimated. However, the model coefficients are not the best way of estimating the overall importance of a feature \cite{PARR2024119563}. This is because the value of each coefficient depends on the scale of the input feature. 

A prediction $\hat{f}(x)$, received by a linear model for one data-point $x$ can be represented as:
\begin{equation}\label{eq1}
\hat{f}(x) =  \beta_0 + \beta_1x_1 + \cdots + \beta_nx_n.
\end{equation}
The $i$-th feature contribution $\phi_i$ to the prediction $\hat{f}(x)$ can be given as:
\begin{equation}\label{eq2}
\phi_i(\hat{f}) = \beta_ix_i -E(\beta_i\mathbf{X}_i),
\end{equation}
where $E(\beta_i\mathbf{X}_i)$ is the average effect estimate for the feature $i$. Considering the sum of all feature contributions for one data-point, it leads to the following:
\begin{equation} \label{eq3}
\begin{aligned}
\sum_{i=1}^{M}\phi_i(\hat{f})  & =  \sum_{i=1}^{M}(\beta_ix_i - E(\beta_i\mathbf{X}_i)) \\
&  =  \left(\beta_0 + \sum_{i=1}^{M}\beta_ix_i\right) -  \left(\beta_0 +\sum_{i=1}^{M}E(\beta_i\mathbf{X}_i)\right) \\
&  =  \hat{f}(x) - E(\hat{f}(\mathbf{X})),
\end{aligned}
\end{equation}
which is the predicted value for the data-point $x$ minus the average predicted value.

\subsection{Model Agnostic Explainablity}
A vector of Shapley values deﬁnes the principle of optimal distribution of gain--predicted value, between the players--features in the theory of cooperative games. It represents a distribution, where the gain of each player is equal to its average contribution to the welfare of the total coalition. Thereby, using the Shapley values, all possible combinations and options are revealed and the features, which are relevant, can be identified.

The contribution of a feature value, e.g., feature $\mathrm{X}3=\mathrm{x}31$, can be demonstrated using a small example, which makes it possible to interpret the meaning of  Shapley value more clearly. For example, the total number of features is 4, and there is a need to determine the contribution of $\mathrm{X3}$ and its value $\mathrm{x}31$.
As a first step, the number of all possible coalitions (combinations) with the 3 remaining features is estimated. This way, $M$ coalitions are formed. If in considered ``coalition-1'' there are, for example, only 2 features, then $\mathrm{X}3=\mathrm{x}31$ is added to this coalition and the missing fourth feature is selected randomly from other samples. A randomly selected feature acts as a `donor'. Then, a prediction $\mathrm{P1}$ of the target value is made. As a next step, the value of $\mathrm{X}3=\mathrm{x}31$ is removed and selected randomly from other instances, and a new prediction $\mathrm{P2}$ is made. The contribution of $\mathrm{X}3=\mathrm{x}31$ is equal to the difference between the predictions made at the first and second iterations. Basically, the prediction values depended on a randomly selected $\mathrm{X}4=\mathrm{x}41$ and further changed value of $\mathrm{X}3=\mathrm{x}32$. To estimate the contribution of $\mathrm{X}3=\mathrm{x}31$ more accurately, it is necessary to repeat this sampling step and then average the contributions received. Thereby, the Shapley value represents the average of all contributions of the feature value in all possible coalitions $M$ (combinations of features). It should be highlighted that Shapley value is not the difference in predictions while removing the feature from the feature vector \cite{molnar2022}.

The Shapley value is a value function $\hat{f}(S)$ of features--players, in the subset--coalition $S$:
\begin{equation} \label{eq4}
\hat{f}(S)=\int_{}^{}\hat{f}(x_S, \mathbf{X}_{\bar{S}} )d\mathrm{P}_{\mathbf{X}_{\bar{S}}} - E_X(\hat{f}(\mathbf{X})),
\end{equation}
where $x_S$ represents the values of the features that are in coalition $S$, $\mathbf{X}_{\bar{S}}$ is the vector of feature values to be explained and considered as complimentary/unobserved features in random variables, i.e., those features that are not in the coalition $S$, and $\mathbf{X}$ is the random variable in the background dataset \cite{molnar2022,meng.svf.ais.2024}. Equation \eqref{eq4} performs multiple integrations for each feature that is not included into the coalition $S$. For example, the model includes four features, i.e., $\mathbf{X}1$, $\mathbf{X}2$, $\mathbf{X}3$, $\mathbf{X}4$ and the coalition considered includes two feature values: $\mathrm{x}{11}$, $\mathrm{x}21$, which leads to the following expression:
\begin{equation} \label{eq5}
\hat{f}(\{1,2\}) = \int_{\mathrm{R}}\int_{\mathrm{R}}\hat{f}(\mathrm{x}11, \mathrm{x}21, \mathbf{X}3,  \mathbf{X}4)\mathrm{d}\mathrm{P}_{\mathrm{X}3\mathrm{X}4} - E_X(\hat{f}(\mathbf{X})).
\end{equation}
Note that equation \eqref{eq5} is similar to the one defined for the linear model, i.e., \eqref{eq3}. Essentially, the evaluation of a feature value occurs by averaging the differences in predictions made when fixing the values of the coalition features and randomly sorting through the values of features that are not included in the coalition. On top of that, all possible combinations of coalitions should be taken into account.

One of the advantages of Shapley values compared to other methods, e.g., \texttt{LIME}, is its efficiency. This means that the difference between the actual prediction and the average prediction is fairly distributed among the feature values:
\begin{equation} \label{eq6}
\sum_{i=1}^{M}\phi_i = \hat{f}(x)-E_X(\hat{f}(\mathbf{X})).
\end{equation}

The axioms of efficiency, symmetry, dummy, and additivity underlying the Shapley values provide a theoretical basis for feature value explainability \cite{molnar2022}. The calculation of Shapley values require high computer resources due to $2^M$ of all possible coalitions of the feature values. This shortcoming, however, can be overcome by computing the contributions for a few samples of the possible coalitions, which reduces the number of iterations but increases the variance of Shapley values. As an alternative, Shapley value can be approximated using the Monte-Carlo sampling.


\section{Modelling}
SHAP (SHapley Additive exPlanations) method splits the forecast into parts to reveal the contribution of each feature \cite{NIPS2017_7062}, and  is based on the Shapley vector. However, it represents the Shapley value explanation as a linear model of coalitions:
\begin{equation}\label{eq7}
g(x^{'}) = \hat{f}(x)=\phi_0 + \sum_{i=1}^{M}\phi_ix_{i}^{'}.
\end{equation}
In case $\phi_0 = E_X(\hat{f}(\mathbf{X}))$ and the coalition vector $x_{i}^{'}$ is set to all ones, meaning that all feature values are present. Then, the Equation \eqref{eq7} turns to a form of Shapley efficiency property \eqref{eq6}:
\begin{equation}\label{eq8}
\hat{f}(x)=\phi_0 + \sum_{i=1}^{M}\phi_ix_{i}^{'} = E_X(\hat{f}(\textbf{X}))+\sum_{i=1}^{M}\phi_i,
\end{equation}
where $\phi_i$  are the Shapley values.
The main difference of SHAP from the classical Shapley values is the coalition vector, that is, if a feature is present in the coalition, then its taken as equal to 1, if it is absent, then 0. Thus, the coalition vector is a vector of zeros and ones. The function $h_x(z^{'})=z$, where $h_x:\{ 0,1\}^{M}\to \mathrm{R^{M}}$ maps a coalition such that for the present features the corresponding/actual value is assigned and for the absent features the values are assigned by random feature values from the data.

\begin{figure}[htbp]
\centering
\includegraphics[width=1.0\columnwidth]{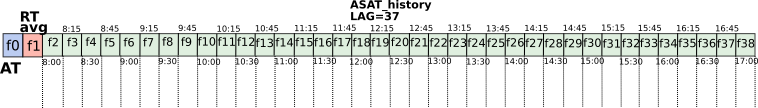}
\caption{Configuration and encoding of the feature vector used in modelling of ASAT.}
\label{Fig3}
\end{figure}

The model predictive control of ASAT employs sliding window approach, which can be interpreted as Iterative Learning Control (ILC), meaning that the next iteration of a system component is predicted using its historical data. This way the learning is implemented via learning from the past \cite{lautenschlager.ddi.ifac.2016}. In the present study the feature vector of a data point included the Ambient Temperature (AT), Room Temperature average (RT-avg) and history of ASAT values.
The configuration and encoding of the feature vector is presented in Fig.~\ref{Fig3}.

\begin{figure}[htbp]
\centering
\includegraphics[width=0.99\columnwidth]{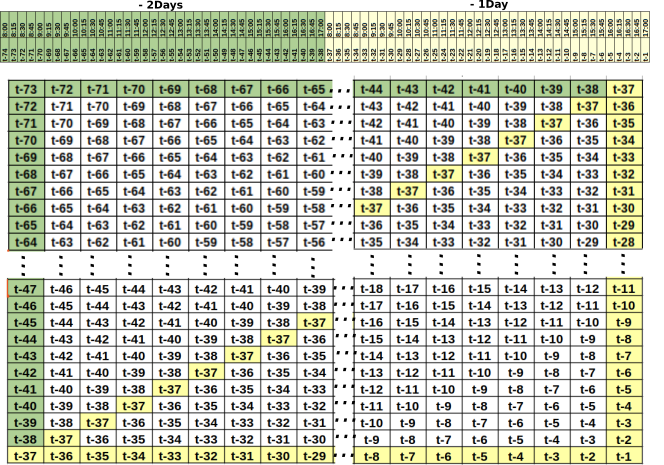}
\caption{Example of a building a 2D Dynamic array of ASAT history based on two previous days. 2D array is given in time-step representation.}
\label{Fig4}
\end{figure}

Building a 2D dynamic array of ASAT history based on two previous days is shown in Fig.~\ref{Fig4}. The lag is equal to 37, which is equivalent to the operational hours of one day starting from 08:00 and until 17:00. The 2D array in Fig.~\ref{Fig4} is given in time-step representation.

The feature vector of ASAT in the first iteration, based on Figs.~\ref{Fig3} and \ref{Fig4}, can be constructed as:
{\scriptsize
\begin{verbatim}
['f3:8:15', 'f4:8:30', 'f5:8:45', 'f6:9:00', 'f7:9:15',
'f8:9:30', 'f9:9:45', 'f10:10:00', 'f11:10:15', 'f12:10:30',
'f13:10:45', 'f14:11:00', 'f15:11:15', 'f16:11:30',
'f17:11:45', 'f18:12:00', 'f19:12:15', 'f20:12:30',
'f21:12:45', 'f22:13:00', 'f23:13:15', 'f24:13.30',
'f25:13:45', 'f26:14:00', 'f27:14:15', 'f28:14:30',
'f29:14:45', 'f30:15:00', 'f31:15:15', 'f32:15:30',
'f33:15:45', 'f34:16:00', 'f35:16:15', 'f36:16:30',
'f37:16:45', 'f38:17:00', 'f2:8:00']
\end{verbatim}}
Feature vectors in other iterations follow the time-step pattern depicted in Fig.~\ref{Fig4}.

\section{Numerical Results and Analysis}
\begin{figure}[htbp]
\centering
\includegraphics[width=0.85\columnwidth]{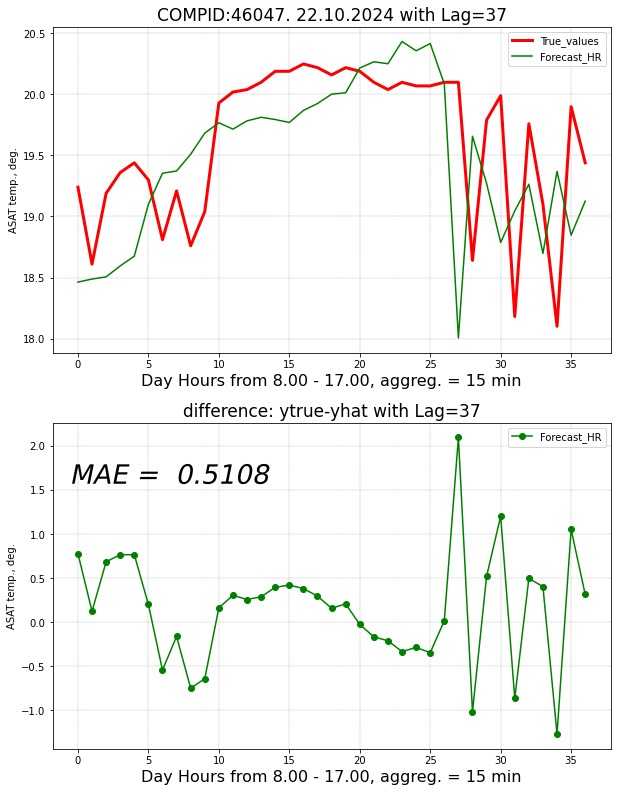}
\caption{The forecasted and true ASAT control values, as well as the difference between the control curves.}
\label{Fig5}
\end{figure}

Explainability of a control curve of ASAT was made based on the forecast for the October 10, 2024, which was a working day.
The forecasted and true ASAT control values, as well as the difference between the control curves are shown in Fig.~\ref{Fig5}. The maximum difference between the true ($y=20.1$) and forecasted ($\hat{y}=18.00$) values of ASAT occurred at 14:45 and was equal to 2.1 [$^{\circ}$C].
Reviewing the distributions of model parameters of ASAT, the following highlights can be marked:
\begin{itemize}
\item Relevant features (referring to the absolute order of magnitude) appear in the beginning and end of the feature vector, indicating the relevance of: AverageRoomTemperature and ASAT historic data corresponding to the end of the working day (only operational hours of AHU were considered while building the model, i.e., 8:00--17:00);
\item An interesting observation is the lower effect of AmbientTemperature.
\end{itemize}

\begin{figure}[htbp]
\centering
\includegraphics[width=0.9\columnwidth]{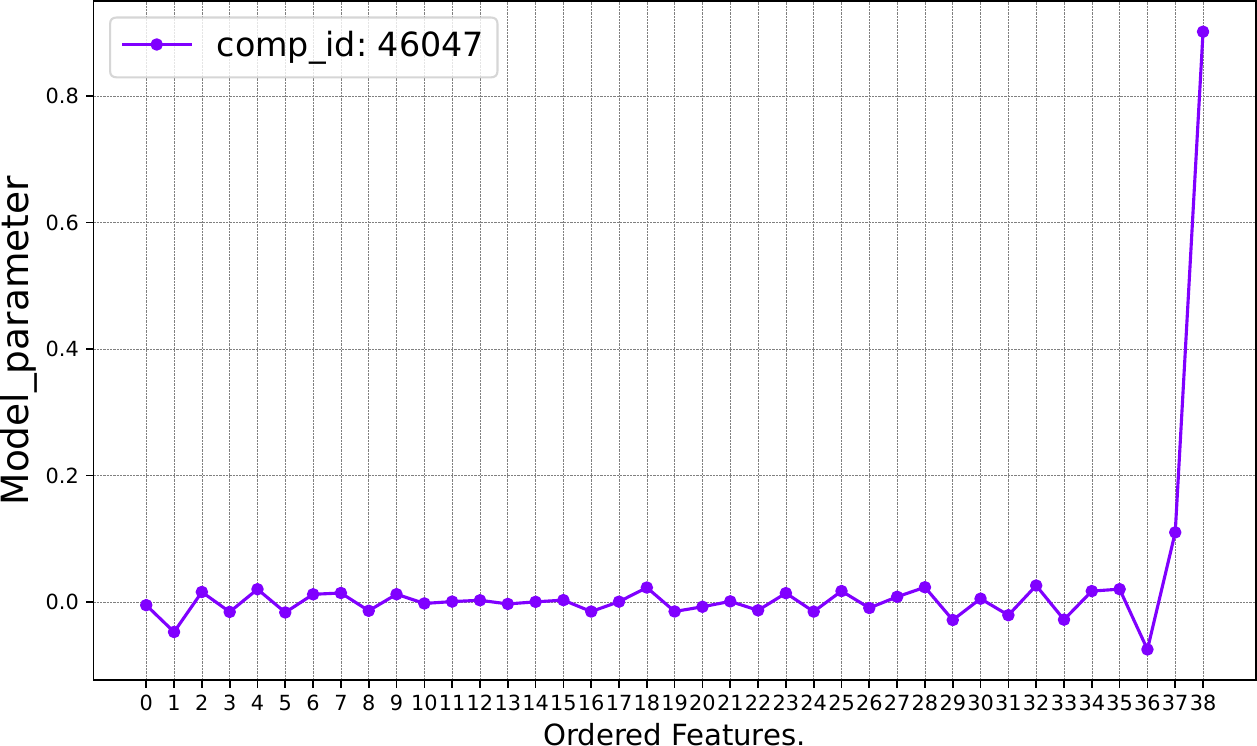}
\caption{Distribution of model parameters--coefficients, of models trained to forecast ASAT control values. ASAT history data in time-step representation.}
\label{Fig6}
\end{figure}
\begin{figure}[htbp]
\centering
\includegraphics[width=0.99\columnwidth]{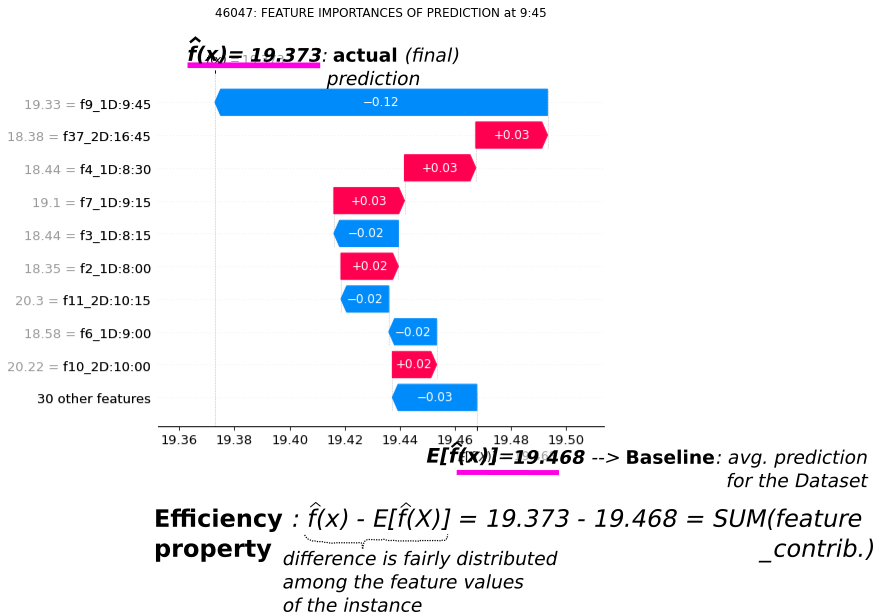}
\caption{Feature importances of forecasted ASAT value at 9:45.}
\label{Fig7}
\end{figure}

Easy explanation and visualisation of ASAT values forecasted considering the 2D dynamic array of features can be made by \texttt{shap.plots.waterfall()}, see \cite{NIPS2017_7062}. The number of such explanations and visualisation slices is equal to 37, i.e., daytime hours from 08:00--17:00 with a resolution 15 min. Figure~\ref{Fig7} demonstrates the importance of the features for forecasting the ASAT value at 9:45. 
The feature importance given in Fig.~\ref{Fig7} using the semantic explanation can be given in the form represented in Table~\ref{tab:Tab1}.
\begin{table}[htbp]
\centering
\caption{Feature Importance Corresponding to Two abs.max Shapley Values using the Semantic Explanation}
\label{tab:Tab1}
\begin{tabular}{C{2.5cm}C{2.5cm}C{2.5cm}}
\toprule
forecast timestamp &
feature-1 abs.max Shapley val. &
feature-2 abs.max Shapley val. \\
\midrule
`8:00'  & `f38\_2D:17:00' & `f2\_1D:8:00'   \\
`8:15'  & `f2\_1D:8:00'   & `f3\_1D:8:15'   \\
`8:30'  & `f3\_1D:8:15'   & `f4\_1D:8:30'   \\
`8:45'  & `f4\_1D:8:30'   & `f5\_1D:8:45'   \\
`9:00'  & `f5\_1D:8:45'   & `f6\_1D:9:00'   \\
`9:15'  & `f6\_1D:9:00'   & `f7\_1D:9:15'   \\
`9:30'  & `f6\_1D:9:00'   & `f8\_1D:9:30'   \\
`9:45'  & `f37\_2D:16:45' & `f9\_1D:9:45'   \\
`10:00' & `f34\_2D:16:00' & `f10\_1D:10:00' \\
`10:15' & `f35\_2D:16:15' & `f11\_1D:10:15' \\
\bottomrule
\end{tabular}
\end{table}

\begin{figure}[htbp]
\centering
\includegraphics[width=0.99\columnwidth]{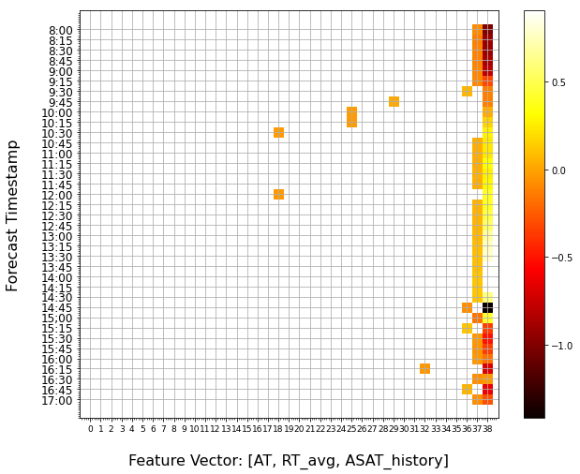}
\caption{Binary representation of the first two features having abs.max Shapley Values per timestamp of the forecast.}
\label{Fig8}
\end{figure}

An interpretation of, for example, a value forecasted for the daytime at 9:45, having the relevant features, such as: `f37\_2D:16:45', `f9\_1D:9:45' can be as: those are the thirty-seventh and ninth features in the ordered feature vector (counting starts from `0'), which correspond to ASAT historical values of the timestamps at `16:45 minus 2-DAYS' and  `9:45 minus 1-DAY', respectively. Using the semanitic explanation given above, it can be represented as: [`9:45', (`f37-2D:16:45', `f9-1D:9:45')]. The possible explanation for the matter of maximum difference maybe be as: the feature that played the most significant role for the forecast was equal to ASAT temperature of the previous day at the same and 15 min earlier timestamps.  A binary visualisation of first two features having abs.max Shapley values per timestamp of the forecast is given in Fig.~\ref{Fig8}. Figure~\ref{Fig8} should be analysed together with Fig.~\ref{Fig4} and semantic explanation represented in Table~\ref{tab:Tab1}, since the feature vector has a dynamic structure.



\section{Conclusions}
The study presented answered the following questions:
\begin{itemize}
\item Forecast explanations produced by \verb@shap.plots.waterfall()@ made it possible to receive the local--contrastive explanations, of forecasts: slices. This added a clarity to each forecast point of the ASAT curve. Besides, the slicing explanations made it possible to justify from the physical and semantic point of views any control curve changes by making this functionality as a customer friendly service.
    \item In comparison with \verb@LIME@, \verb@SHAP@, i.e. application of Shapley values, always use all the features to explain the prediction. The efficiency property of Shapley value allows to estimate the difference between the final and the average predictions, which is fairly distributed among the contributions of the feature values.
\verb@LIME@, in turn, does not allow such transparency. 
\item  \verb@SHAP@ enabled for global model interpretations with all features always present, whereas in the case of \verb@LIME@ only local explanations are possible.
\end{itemize}


\bibliographystyle{IEEEtran}
\bibliography{Bibliography.bib}

\end{document}